\documentclass[universe,article,accept,pdftex,moreauthors]{Definitions/mdpi}
\firstpage{1} 
\pubvolume{1}
\issuenum{1}
\articlenumber{0}
\pubyear{2024}
\copyrightyear{2024}
\datereceived{ } 
\daterevised{ } 
\dateaccepted{ } 
\datepublished{ } 
\hreflink{https://doi.org/} 

\Title{A phenomenological model of the magnetic field re-emergence in magnetars and discrepancy between the kinematic and characteristic ages}

\TitleCitation{Field re-emergence and magnetars ages}

\Author{Rostislav D. Nikandrov $^{1}$*\orcidA{} and Sergei B. Popov $^{2}$*\orcidB{}}

\AuthorNames{Rostislav Nikandrov and Sergei B. Popov}

\AuthorCitation{Nikandrov R.D.; Popov, S.B.}

\address{%
$^{1}$ \quad Department of Physics, Lomonosov Moscow State University, 1/2 Leninskie Gory, 119991, Moscow, Russia\\
$^{2}$ \quad Sternberg Astronomical Institute, Lomonosov Moscow State University, Universitetskij pr. 13, 119234, Moscow, Russia 
}

\corres{Correspondence: sergepolar@gmail.com; Tel.:  +7 495 9395006}

\abstract{ Robust age measurements for isolated neutron stars (NSs) are not easily available. That is why, often the characteristic age $\tau_\mathrm{ch}=P/2\dot P$ is used as a proxy. Here $P$ is the spin period of the NS and $\dot P$ is the time derivative of $P$. Additional assumptions related to the initial properties and spin-down evolution are made to derive $\tau_\mathrm{ch}$. As a result, it is expected that $\tau_\mathrm{ch}$ is an upper limit for the real age $\tau_\mathrm{real}$. Recently, Chrimes et al. presented measurements of kinematic ages $\tau_\mathrm{kin}$ for several magnetars. Surprisingly, for the majority of these sources $\tau_\mathrm{kin}>\tau_\mathrm{ch}$. We present a simple model including a realistic approximation for the magnetic field decay in magnetars and a simple phenomenological description of the field re-emergence after an episode of fallback after the birth of a NS. We demonstrate that this simple model can explain the observed relation $\tau_\mathrm{kin}>\tau_\mathrm{ch}$ for realistic sets of parameters. }


\begin{document}

\section{Introduction}

Young neutron stars (NSs) manifest themselves as sources with diverse properties that reflect a variety of their initial properties and evolutionary paths \cite{2023Univ....9..273P, 2023hxga.book..146B, 2024mbhe.confE..55R}. In particular, non-trivial behavior can be a result of initial fallback \cite{1989ApJ...346..847C} and further magnetic field evolution \cite{2021Univ....7..351I}.

 The magnetic field evolution is especially important for magnetars: anomalous X-ray Pulsars (AXPs) and Soft $\gamma$-ray Repeaters (SGRs) \cite{2015RPPh...78k6901T}. 
 Recently, Chrimes et al. \cite{2026arXiv260315750C} used new observations with the {\it Hubble } Space Telescope and {\it James Webb} Space Telescope to measure proper motions of several magnetars. The measurements allowed them to determine the kinematic ages of these objects. Altogether, Chrimes et al. \cite{2026arXiv260315750C} present kinematic age estimates for nine magnetars. Interestingly, for eight objects in the sample, the kinematic ages, $\tau_\mathrm{kin}$, are greater than their characteristic ages $\tau_\mathrm{ch}\equiv P/2\dot P$. Here $P$ is the spin period of a NS and $\dot P$ --- its first time derivative. 

 If correctly determined, a kinematic age represents the actual age ($\tau_\mathrm{real}$) of the NS. On the other hand, for the standard evolution according to the magneto-dipole formula with a constant effective dipolar magnetic field, $\tau_\mathrm{ch}$ is an upper limit for $\tau_\mathrm{real}$ as the initial spin period $P_0$ can be not much smaller than the present-day period. If the magnetic field significantly decays, as is usually assumed in the case of magnetars, we also expect $\tau_\mathrm{ch}\gg \tau_\mathrm{real}$ even for $P_0\ll P$. 

 The inequality $\tau_\mathrm{ch} < \tau_\mathrm{real}$ indicates that the effective external dipolar magnetic field, which is responsible for the spin-down, is growing. Potentially, this can occur due to changes (growth) of the angle $\chi$ between the spin axis and the magnetic dipole axes \cite{2025ApJ...995..167B}. Or, due to the dynamics of the field in the interiors of the NS \cite{2020MNRAS.495.1692G, 2025A&A...694A..39D, 2023hxga.book..146B}. Finally, the external field, which determines the spin-down of the NS, can increase due to re-emergence after an initial episode of the fallback \cite{2011MNRAS.414.2567H, 2012MNRAS.425.2487V, 2016MNRAS.462.3689I}.

 In this paper, we interpret the result by Chrimes et al. \cite{2026arXiv260315750C} in the framework of a re-emerging magnetic field accounting for the field decay. For the rate of re-emergence, we suggest a simple phenomenological model, and for the decay, we follow the approximation suggested for magnetars in \cite{2008A&A...486..255A}. 

 In the next section, we describe our approach. Then in Sec.~3, we present our results. 
 This is followed by a Discussion. Our conclusions are given in the final section. 
 
\section{Model}
\label{sec_model}

In this study, the spin-down of an NS is calculated following a very basic approach based on a modified spin-down formula:

\begin{equation}
    P \dot P = \frac{4 \pi^2B^2R^6}{Ic^3} (k_0+k_1\sin^2 \chi).
    \label{ppdot}
\end{equation}
Here, $R=10^6$~cm is the NS radius, $B(t)$ is the surface equatorial magnetic field, $I=10^{45}$~g~cm$^2$ is the moment of inertia. 
The values of the coefficients are taken as: $k_0=1$, $k_1=1.2$ \cite{2014MNRAS.441.1879P}. 
The magnetic obliquity angle is assumed to be constant $\chi=45^\circ$.

As we want to apply a model for the re-emerging magnetic field, we need to specify $B(t)$. 
Instead of direct calculations of the external field growth \cite{2011MNRAS.414.2567H, 2012MNRAS.425.2487V, 2016MNRAS.462.3689I}, we suggest a simple equation, which potentially can represent the field growth due to different reasons, not only re-emergence after a fallback episode (however, below we focus only on this possibility):

\begin{equation}
B(t) = B_\mathrm{i}+ (B_\mathrm{c}(t)-B_\mathrm{i}) \times \exp(-t_\mathrm{cr}/t). 
\label{b_evol}
\end{equation}
Here, $B_\mathrm{c}(t)$ is the value of the field at the bottom of the crust. Initially, it is equal to the total buried field $B_\mathrm{max}$. Without field decay, this is the maximum value that can be reached. $B_\mathrm{i}$ is the initial external field; i.e., this is the field after the fallback episode is over. In our calculations below, we typically assume $B_\mathrm{i}=10^{9}$~G and $B_\mathrm{max}=10^{15}$~G , as we are interested in the evolution of magnetars. Finally, $t_\mathrm{cr}$ is the characteristic time scale of field growth.

\begin{figure}[t]
\centering
\includegraphics[width=\textwidth]{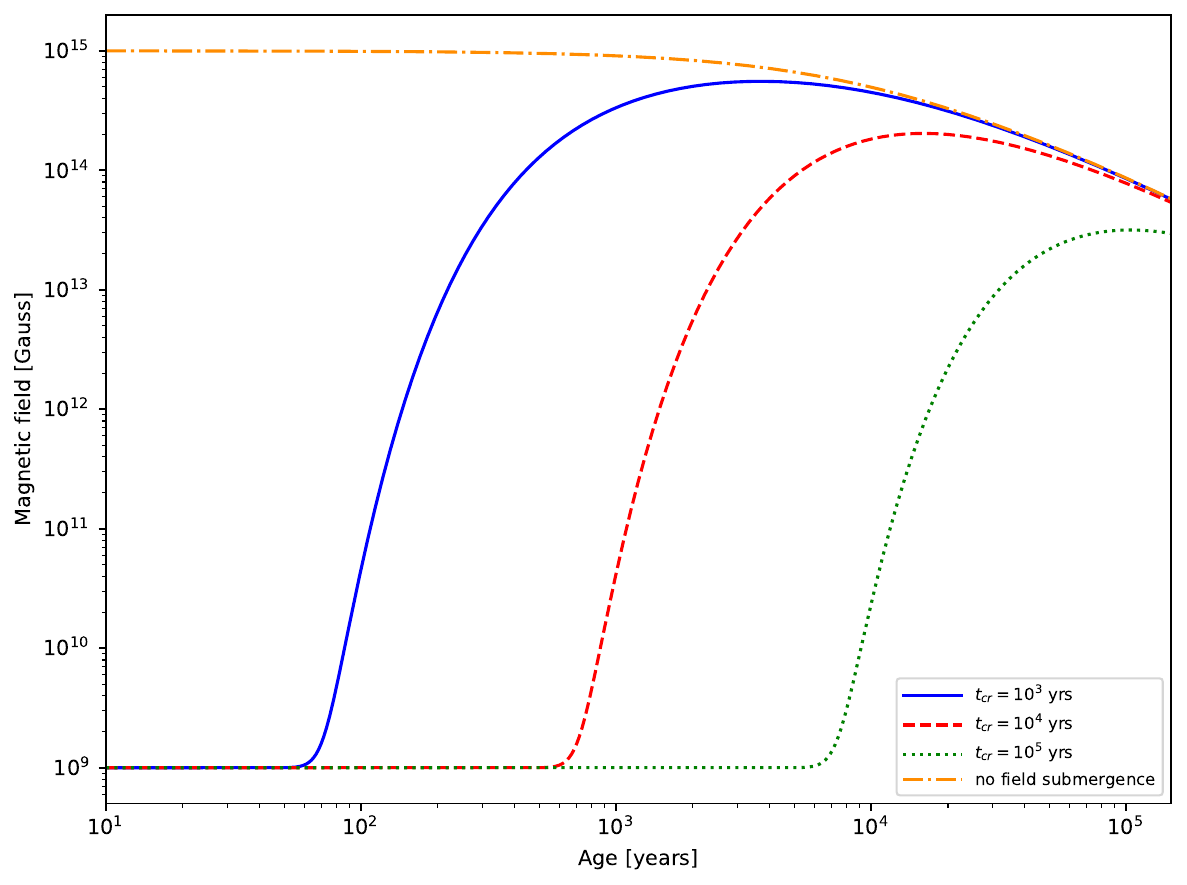}
\caption{Evolution of the external magnetic field $B(t)$. Blue solid line corresponds to $t_\mathrm{cr}=10^3$~yrs, red dashed line --- to $t_\mathrm{cr}=10^4$~yrs, and green dotted line --- to $t_\mathrm{cr}=10^5$~yrs. Orange dash-dotted line represents the model without field submergence for which $B(t)=B_\mathrm{c}(t)$. 
}
\label{f:b_evol}
\end{figure}

To take into account the magnetic field decay, we apply an equation from \cite{2008A&A...486..255A}:

\begin{equation}
B_\mathrm{c}(t)= B_\mathrm{max} \frac{\exp(-t/\tau_\mathrm{Ohm})}{1+(\tau_\mathrm{Ohm}/\tau_\mathrm{Hall})(1-\exp(-t/\tau_\mathrm{Ohm})}.
\label{b_decay}
\end{equation}
Here, $\tau_\mathrm{Ohm}$ is the Ohmic time scale. And $\tau_\mathrm{Hall}$ is the Hall time scale. In the calculations, we typically assumed that $\tau_\mathrm{Ohm}=10^6$~yrs and $\tau_\mathrm{hall}=10^4$~yrs for $B_\mathrm{c}(t=0)=B_\mathrm{max}$. The magnetic field evolution for several parameter values is shown in Fig.~\ref{f:b_evol}. Note that decay is regulated by the value of the decaying field in the crust, $B_\mathrm{c}$. This narrows the applicability of our model only to the case of the field buried in a fallback. For other models of the growth of the effective field, it is necessary to modify the equations. 

In our calculations, we do not take into account the Hall attractor \cite{2014PhRvL.112q1101G}. This stage is expected to occur after $\approx 3$ $e$-foldings of the field decay \cite{2014MNRAS.438.1618G}. 
Substituting this condition into eq.~(\ref{b_decay}), we obtain the characteristic time of transition to the Hall attractor $t_\mathrm{HA}=1.7\times 10^5$~yrs for the initial field $B_\mathrm{max}=10^{15}$~G. 
This value exceeds the typical age of the magnetars under study. But, if we take smaller values of $\tau_\mathrm{Hall}$, accounting for the Hall attractor becomes necessary. We discuss this in detail in Sec.~\ref{sec_disc}.


\section{Results}
\label{sec_res}

\begin{figure}[t]
\centering
\includegraphics[width=\textwidth]{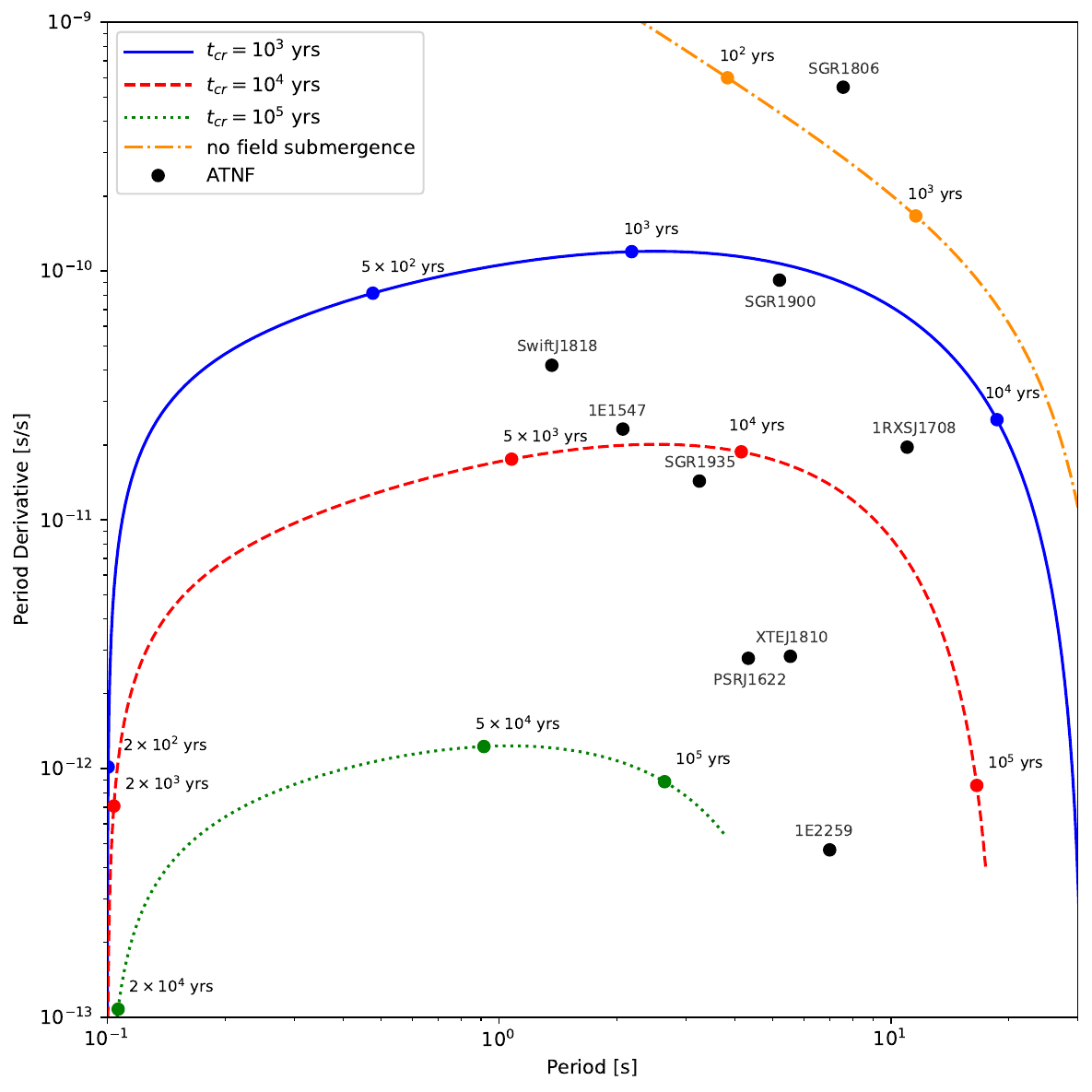}
\caption{$P$--$\dot P$ diagram. Blue solid line corresponds to $t_\mathrm{cr}=10^3$~yrs, red dashed line --- to $t_\mathrm{cr}=10^4$~yrs, green dotted line --- to $t_\mathrm{cr}=10^5$~yrs, and orange dash-dotted line --- to the model without field submergence. On each curve, we put time-ticks.
Data points are taken from the ATNF pulsar catalogue \cite{2005AJ....129.1993M}.
}
\label{f:ppdot}
\end{figure}

In this section, we present the results of our calculations. 
In all cases, we assume that the initial field before submergence is $B_\mathrm{max}=10^{15}$~G and just after submergence $B_\mathrm{i}=10^9$~G. The initial spin period is always taken as $P_0=0.1$~s in correspondence with the proposed distributions \cite{2012Ap&SS.341..457P, 2022MNRAS.514.4606I} (note that the results do not depend significantly on this value since $P_0\ll$~a few seconds).
We use three values of the field re-emergence time scale $t_\mathrm{cr}$: $10^3, \, 10^4 $, and $10^5$~yrs. The middle value approximately corresponds to the calculations in \cite{2011MNRAS.414.2567H, 2012MNRAS.425.2487V}. The longest time scale $t_\mathrm{cr}=10^5$~yrs is of the order of the time scale obtained in \cite{2016MNRAS.462.3689I} (however, these authors modeled a pulsar-like field). Obviously, for a smaller amount of fallback matter, the characteristic time scale of re-emergence is shorter. Thus, we also present results for $t_\mathrm{cr}=10^3$~yrs and for the absence of the field submergence. 

We numerically solve eq.~(\ref{ppdot}), accounting for the field evolution according to the eqs.~(\ref{b_evol}) and (\ref{b_decay}). Spin period and period derivative evolution for different values of $t_\mathrm{cr}$ is presented in the $P$--$\dot P$ diagram (Fig.~\ref{f:ppdot}). 


Each curve for a submerged field starts at $\dot P=2.3\times 10^{-20}$. The early evolution with rapid growth of $\dot P$ is not shown. The curves for $t_\mathrm{cr}=10^4$ and $10^5$~yrs are truncated at $\approx 1.5\times 10^5$~yrs before the Hall attractor is reached for the chosen value of $\tau_\mathrm{Hall} $ (see Sec.~\ref{sec_disc}).  Three stages can be distinguished. In the first brief phase of evolution, $\dot P$ and $B$ rapidly grow while $P$ is not increasing significantly. Then the magnetic field continues to increase untill the maximum value is reached, and the NS spins down significantly. Finally, the $P$~--~$\dot P$ evolution is dominated by the field decay.

\begin{figure}[t]
\centering
\includegraphics[width=\textwidth]{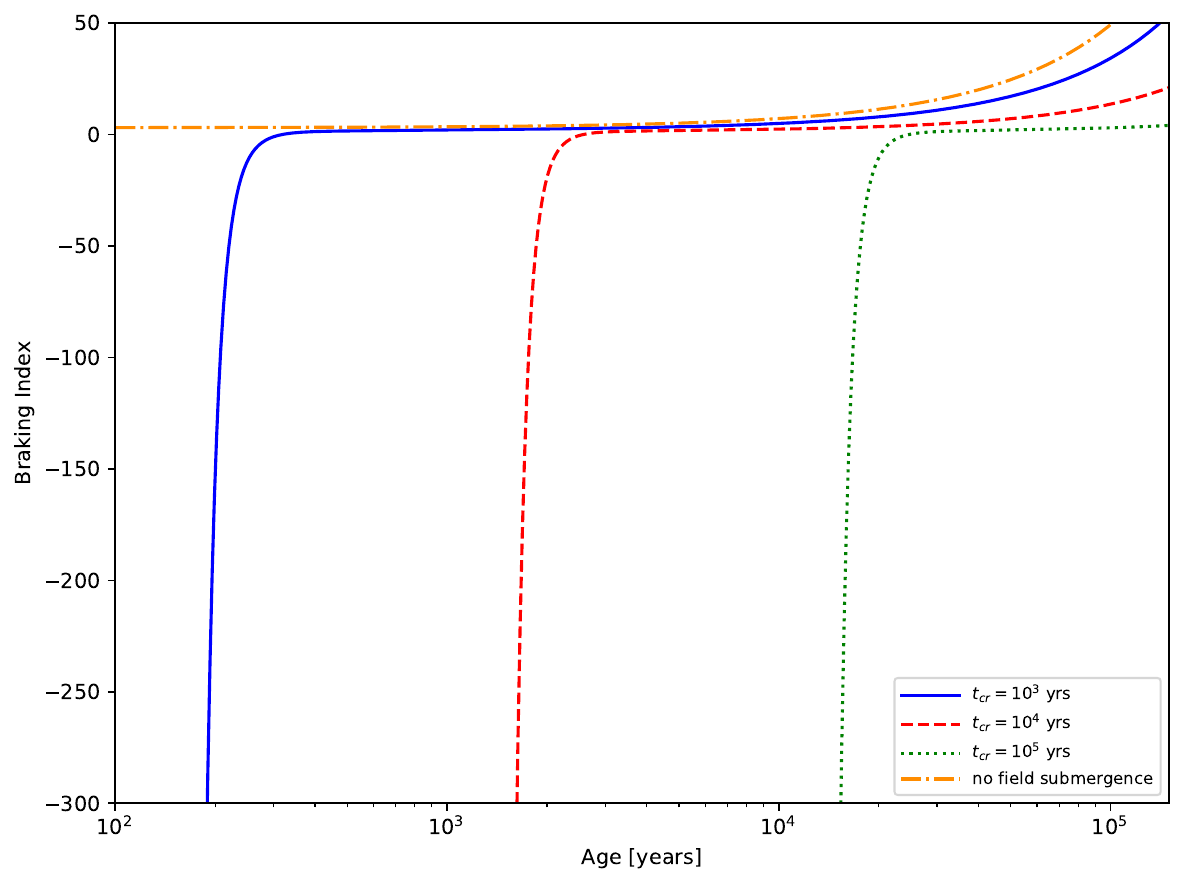}
\caption{Evolution of the braking index. Blue solid line corresponds to $t_\mathrm{cr}=10^3$~yrs, red dashed line --- to $t_\mathrm{cr}=10^4$~yrs, green dotted line --- to $t_\mathrm{cr}=10^5$~yrs, and orange dash-dotted line --- to the model without field submergence.
}
\label{f:bindex}
\end{figure}

To illustrate the evolution of an NS during the field re-emergence, we also plot the behavior of the braking index $n$ (Fig.~\ref{f:bindex}):
\begin{equation}
    n = {\frac {\nu \ddot \nu} {\dot \nu^{2}}} = 2 - \frac{P\ddot P} {\dot P^{2}},
    \label{bindex}
\end{equation}
where $\nu=1/P$ is the spin frequency.
In this plot, we omit a brief episode when $n\approx3$ at an early stage of evolution when $B\approx B_{i}$ (see Sec.~\ref{sec_disc}). 

In the stage of magnetic field re-emergence, the braking index is negative. It rapidly approaches the classical value $n=3$. In the later stages, the field evolution is dominated by decay. Since decay dominates, the braking index has large positive values. This stage is over when the decay saturates, e.g., when the NS reaches the Hall attractor stage.  

Unfortunately, due to the noisy behavior of magnetars, precise measurements of the second spin period derivative are not possible. Still, potentially it is possible to distinguish the sources at the field re-emergence stage by their large negative braking indices. E.g., according to the ATNF catalogue \cite{2005AJ....129.1993M}, several magnetars from the sample we use have large negative braking indices:
RXS J170849 and SGR 1806-20.
Several others have large positive braking indices: PSR J1622 has $n=1.1\times 10^4$, XTE J1810 has $n=870$, and SGR 1935 has $n=5.5\times 10^3$ (see Table 1 and discussion below).

\begin{figure}[t]
\centering
\includegraphics[width=0.9\textwidth]{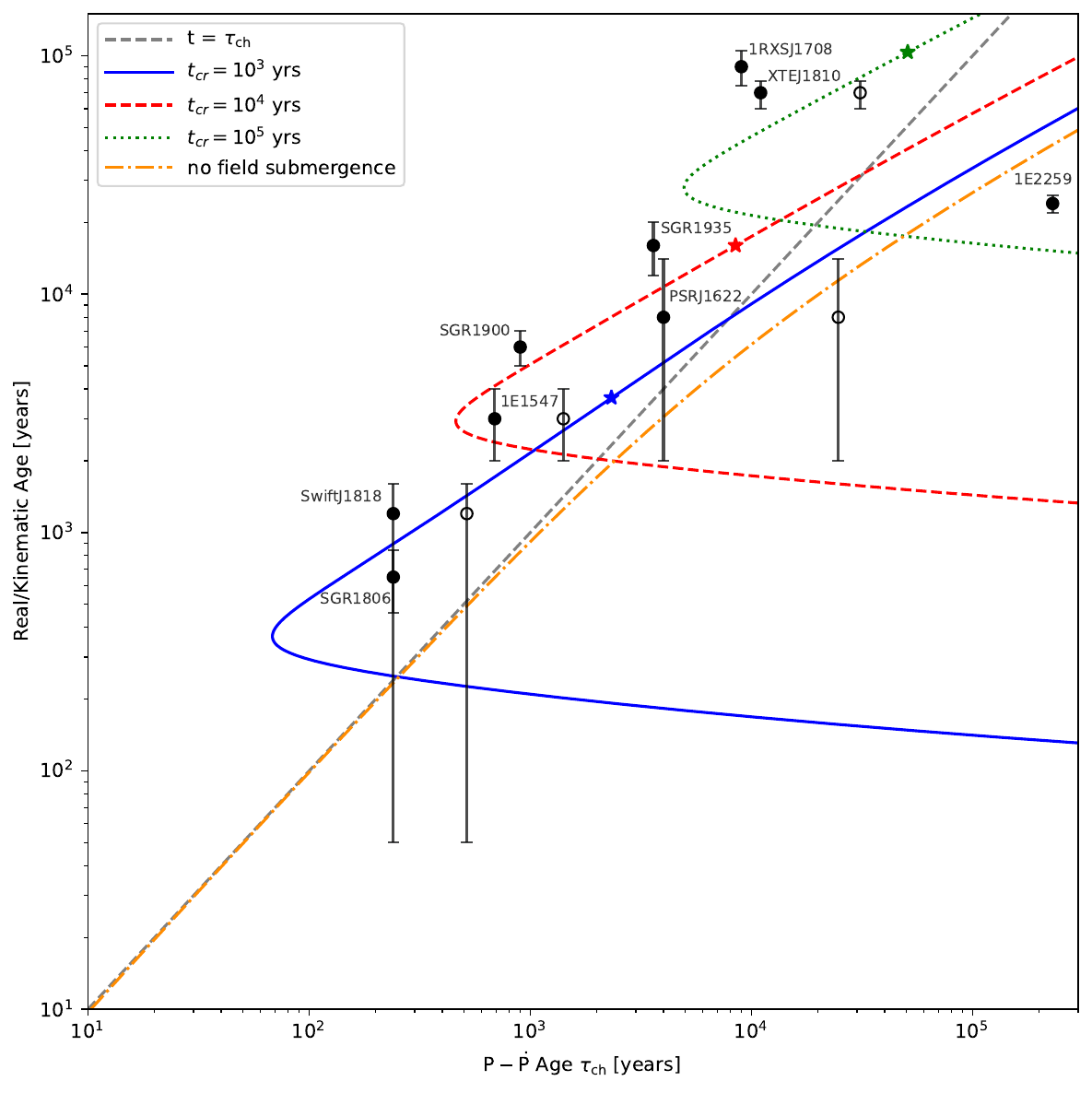}
\caption{Real/kinematic ages vs. characteristic ages of magnetars. 
Data points shown with filled black circles 
are taken from Chrimes et al. \cite{2026arXiv260315750C}. 
If $\tau_\mathrm{ch}$ in \cite{2026arXiv260315750C} and ATNF do not match, we add a second point (open symbols) for each such source with $\tau_\mathrm{ch}$ from the catalogue. 
The colored lines illustrate the evolution of the modeled NSs.
Blue solid line corresponds to $t_\mathrm{cr}=10^3$~yrs, red dashed line --- to $t_\mathrm{cr}=10^4$~yrs, green dotted line --- to $t_\mathrm{cr}=10^5$~yrs, and orange dash-dotted line --- to the model without field submergence. On each curve, we mark with a star symbol the moment when the maximum value of the external dipolar field is reached. 
Grey dashed bisector is plotted for $\tau_\mathrm{ch}=\tau_\mathrm{kin}.$
}
\label{f:age_age}
\end{figure}

Our main results are presented in Fig.~\ref{f:age_age}. We plot $\tau_\mathrm{kin}$ vs. $\tau_\mathrm{ch}$ for nine magnetars from \cite{2026arXiv260315750C}. For four notable objects (1E 1547, PSR J1622, XTE J1810, and Swift J1818) estimates of $\tau_\mathrm{ch}$ in \cite{2026arXiv260315750C} and the ATNF catalogue are significantly (a factor $\gtrsim 2$ different. In such cases, we plot the second point for each of the four sources (see also discussion in Sec.~\ref{sec_disc}). 
We overplot four curves based on our model of the magnetic field evolution. Three curves correspond to different values of $t_\mathrm{cr}$ --- $10^3, \, 10^4,$ and $10^5$~yrs. The fourth line is plotted for the non-submerged field evolution. 
To guide an eye, we also show the bisector $\tau_\mathrm{real}=\tau_\mathrm{ch}$.  

 The curves for the re-emerging field can be described by three stages (they are also visible in Fig.~\ref{f:ppdot}, see description above). In the first one (small $\tau_\mathrm{real}$ and initially large $\tau_\mathrm{ch}$, which rapidly decreases), the external field is significantly suppressed so that $B(t)\approx B_i$ (compare with Fig.~\ref{f:b_evol}). At this stage, the evolutionary curve crosses the bisector.  In the second stage, the external field rapidly grows and reaches its maximum value. At this stage,  $\tau_\mathrm{real}$ > $\tau_\mathrm{ch}$. Finally, in the third stage, the field evolution is governed by decay. The curves cross the bisector back. Thus, finally we have
 $\tau_\mathrm{real}$ < $\tau_\mathrm{ch}$.

 We see that the curves calculated for the re-emerging field can naturally explain the data points with $\tau_\mathrm{kin}$ > $\tau_\mathrm{ch}$. On the other hand, the source AXP 1E 2259+586 can be explained either by a decaying field alone or by the stage of evolution in which the re-emergence occurred some time ago, with the NS properties regulated mainly by the field decay. 
 

\section{Discussion}
\label{sec_disc}

 As our model is not based on detailed calculations of the field diffusion, its applications are limited. 
In particular, we do not attempt to fit the parameters of individual sources (magnetic field, amount of fallback $\Delta M$, etc.). Some objects can be at the stage of field re-emergence. In others, this stage can be over already, or the fallback was negligible and the field had not been submerged. Our main goal is to demonstrate that for a natural choice of parameters, it is possible to explain the situation where $\tau_\mathrm{ch}< \tau_\mathrm{real}$. 

At least for one source in the sample from \cite{2026arXiv260315750C} --- AXP 1E 2259+586 --- it is observed that $\tau_\mathrm{ch}> \tau_\mathrm{real}$. Obviously, this source can be explained naturally without invoking magnetic field re-emergence. However, it is necessary to note that characteristic ages of magnetars are a subject of significant uncertainty, as the first spin period derivative of these sources is a subject of strong fluctuations. E.g., $\tau_\mathrm{ch}$ of PSR J1622 presented in   \cite{2026arXiv260315750C} is taken from McGill catalogue \cite{2014ApJS..212....6O}. It is based on $\dot P$ measurements from 
\cite{2012ApJ...751...53A}. 
However, in the ATNF catalogue the period derivative is an order of magnitude smaller
\cite{scs+17}. 
If we use this value, then for PSR J1622 we obtain  $\tau_\mathrm{ch}> \tau_\mathrm{kin}$ (see Fig.~\ref{f:age_age}). 

We also face some difficulties in interpreting SGR 1806-20 as it requires extreme parameters and some fine tuning. On the one hand, for this source $\tau_\mathrm{ch} < \tau_\mathrm{kin}$ and it has a moderately high negative braking index $n=-25.7$. On the other hand, it has a large magnetic field ($>10^{15}$~G) according to the ATNF catalogue \cite{2005AJ....129.1993M}, which is hard to find in correspondence with the evolution with field submergence and subsequent re-emergence. We speculate that the reason may be related to the application of eq.~(\ref{ppdot}) to calculate the characteristic age.  
Spin-down of magnetars can significantly deviate from the usual equation due to winds, glitches, anti-glitches, and other factors, e.g. \cite{2015PhRvD..91f3007H, 2015RPPh...78k6901T}.
 Thus, even accounting for the decaying magnetic field, age estimates based on a momentary application of eq.~(\ref{ppdot}) can provide a wrong value for magnetars. The characteristic age is larger for a given $P$ and $\dot P$ if, for the constant field and magnetic obliquity, $n<3$. E.g., braking indices $<3$ are quite typical for young, well-studied pulsars, see e.g. \cite{2002ASPC..271....3K, 2007AdSpR..40.1491Y}. Otherwise, the second (and maybe the first) derivative are not precisely know to get a correct estimate for $\tau_\mathrm{ch}$ and $n$. 
However, the properties of SGR 1806-20 can be fitted in the framework of the model developed in this note. To do so, it is necessary to assume a large submerged field ($B_\mathrm{max}$) and rapid re-emergence. However, as we do not intend to fit individual sources in detail, this discussion is beyond the scope of this paper.

\begin{table}[t]
  \renewcommand{\tabularxcolumn}[1]{>{\centering\arraybackslash}m{#1}}
  \centering
  \caption{Magnetars with measured $\ddot P$ according to the ATNF catalogue. }
  \label{tab:ddotp}
  \setlength{\extrarowheight}{5pt}
\begin{tabular}{|l|c|c|c|c|c|c|}
    \hline
    Name & $P$ & $\dot P$  & $\ddot P$  & $n$ & $\tau_\mathrm{ch}$  & $\tau_\mathrm{kin}$ \\ 
    & [s] & [$10^{-12}\,$s/s] & [$10^{-21}\, $s/s$^{2}$] &  & [kyr] & [kyr] \\ \hline
    SGR 1833-0832 & $7.57$ & $3.4$ & $7.4\times 10^{2}$ & $-4.8\times 10^{5}$ & $34.9$ & - \\
    \hline
    SGR J1745-29 & $3.76$ & $17.6$ & $1.5\times 10^{3}$ & $-1.8\times 10^{4}$ & $3.4$ & - \\
    \hline
    1E 1841-045 & $11.79$ & $40.9$ & $54$& $-380$ & $4.6$ & - \\
    \hline
    RXS J170849.0- & $11.01$  & $19.6$  & $12$ & $-350$ & $8.9$  & $90\pm{15}$ \\
    400910 & & & & & & \\ \hline
    SGR 1806-20 & $7.56$ & $549$ & $1.1\times 10^{3}$ & $-26$ & $0.22$ &  $0.65\pm{0.19}$\\
    \hline
    XTE J1810-197 & $5.54$ & $2.83$ & $-1.3$ & $870$ & $31$  & $70^{+8}_{-10}$ \\
    \hline
    SGR 1935+2154 & $3.25$ & $14.3$ & $-3.5\times 10^{2}$ & $5.5\times 10^{3}$ & $3.6$ & $16\pm{4}$ \\
    \hline
    PSR J1622-4950 & $4.33$  & $2.78$  & $-20$ & $1.1\times 10^{4}$  & $24.7$ & $8\pm{6}$ \\
    \hline
    SGR 0501+4516 & $5.76$ & $5.82$ & $-2.3\times 10^{2}$ & $3.9\times 10^{4}$ & $15.7$  & - \\
    \hline
  \end{tabular}
\end{table}

 Data presented in the ATNF catalogue \cite{2005AJ....129.1993M} allow for the calculations of braking indices for nine magnetars (SGRs and AXPs). Their parameters are summarized in Table~\ref{tab:ddotp} (sources are ordered according to their braking indices).
 All data except $\tau_\mathrm{kin}$ are taken from the ATNF catalogue. The kinematic ages (if known) are taken from \cite{2026arXiv260315750C}.
 Five of them are presented among the objects with measured $\tau_\mathrm{kin}$ \cite{2026arXiv260315750C}. Three of the remaining four have high negative values of $n$. However, as the second spin frequency derivative is not well-measured for magnetars, it would be premature to say that these objects are at the stage of the magnetic field re-emergence. 
 Sources XTE J1810 and SGR 1935 with positive braking indices potentially can be explained as NSs with initially submerged fields, which are now in the evolutionary stage dominated by decay (see Fig.~\ref{f:ppdot}).

\begin{figure}[t]
\centering
\includegraphics[width=0.9\textwidth]{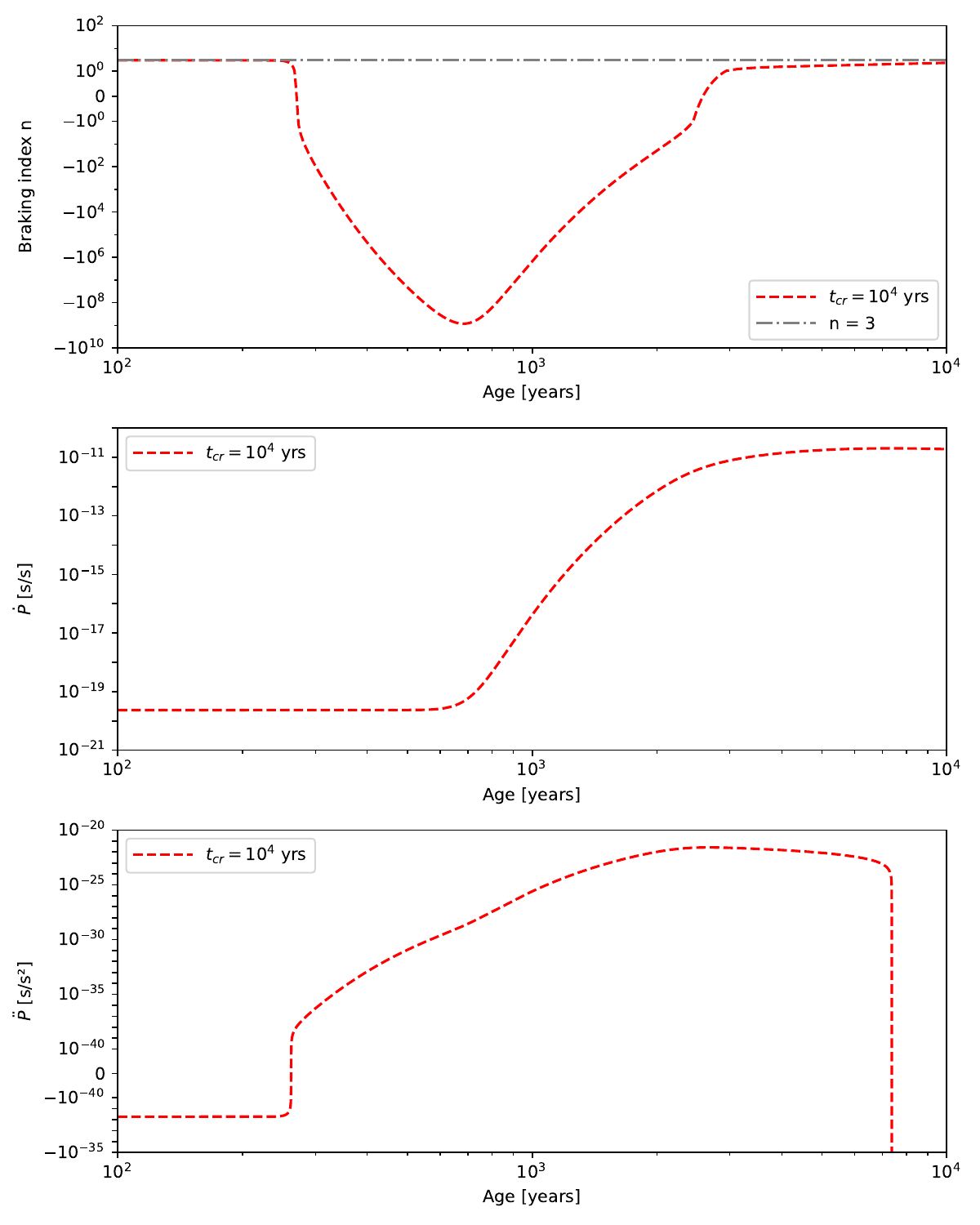}
\caption{Evolution of the braking index, spin period derivative, and the second derivative for $t_\mathrm{cr}=10^4$~yrs.
Top: evolution of $n$. Middle: evolution of $\dot P$. Bottom: evolution of  $\ddot P$.}
\label{f:n_detail}
\end{figure}

In our calculations, we neglect the stage of the Hall attractor. 
As originally proposed in \cite{2014MNRAS.438.1618G, 2014PhRvL.112q1101G}, this stage can be reached after the dipolar field has decayed by a factor $\sim (e^2$~--~$e^3)$. In the case of magnetar-scale initial magnetic field and depending on the value of $\tau_\mathrm{Hall}$, this can happen in a few~$\times 10^3$~--~$10^5$~yrs. 

 We made additional runs of the code, accounting for the Hall attractor stage for the same set of $B_\mathrm{max}, \, B_i, \, \tau_\mathrm{Ohm}$, and the same three values of $t_\mathrm{cr}$. 
 For $\tau_\mathrm{Hall}=10^4$~yrs, as we mentioned above, the transition to the Hall attractor happens 
 at $t=1.7\times 10^5$~yrs. Thus, in this case, the inclusion of this stage does not influence our results. If $\tau_\mathrm{Hall}=10^3$~yrs then the transition happens at $t=1.9\times 10^4$~yrs. 
 This marginally modifies our results, but leaves the main conclusions untouched, as we do not intend to fit all individual sources. Thus, we conclude that as soon as the goal is to explain active magnetars, i.e., the sources which, by definition, did not reach the Hall attractor stage, its inclusion is not necessary. It is necessary only to ensure that the ages of objects in the model do not exceed the transition time for the chosen set of parameters, in particular $B_\mathrm{max}, \, \tau_\mathrm{Hall}, \, t_\mathrm{cr}$. 
 
  
 Finally, we have to make a brief comment on the very early behavior of the braking index in our model. 
 In Fig.~\ref{f:bindex}, we neglected this stage as objects of interest are older than the duration of this early phase. 
The early evolution of $n$, $\dot P$, and $\ddot P$ for $t_\mathrm{cr}=10^4$~yrs is shown in Fig.~\ref{f:n_detail}.
Focusing on the early stages, we truncate this plot at $t=10^4$~yrs. After this time, for the chosen set of parameters, the field evolution is dominated by decay. 

 The evolution of $\dot P$ (middle panel in Fig.~\ref{f:n_detail}) is rather obvious. At first, while $t\ll t_\mathrm{cr}$, the period derivative is small as $B(t)\approx B_i$. Then, the period derivative grows as the field starts to re-emerge. Finally, the growth saturates as $\tau_\mathrm{Hall}$ is of the same value as $t_\mathrm{cr}$. 

 The evolution of the braking index and of the second spin period derivative in the early stage $2\times 10^2\lesssim t \lesssim 7\times 10^2$~yrs, is less trivial (top and bottom panels in Fig.~\ref{f:bindex}). A rapid `jump' in $n$ and $\ddot P$ and the exact shape of the curves representing the decrease of $n$ and the increase of $\ddot P$ are artifacts of the chosen parameterization of the field evolution, i.e., eqs.~(\ref{b_evol}) and (\ref{b_decay}). Thus, this behavior is unphysical. Fortunately, it happens at early ages, which are irrelevant for our discussion.

\section{Conclusions}
\label{sec_concl}

 In this study, we presented a simple model of the evolution of magnetars' external magnetic fields, aiming to explain the relation between kinematic and characteristic ages reported by Chrimes et al. \cite{2026arXiv260315750C}. The key ingredient of our model is a simplified one-parameter description of the magnetic field re-emergence after an episode of fallback. In our framework, we reproduced successfully $\tau_\mathrm{kin}>\tau_\mathrm{ch}$ under the assumption that the magnetic field of some magnetars has been significantly submerged due to fallback. The high negative braking indices reported for several magnetars may indicate that they are still in the stage of field re-emergence.    

\vspace{6pt}
\authorcontributions{The authors contributed to this study equally. All authors have read and agreed to the published version of the manuscript.
}

\acknowledgments{We thank Drs. Anton Biryukov and  Andrei Igoshev for discussions. }

\funding{ SP acknowledges support from the RSF grant 25-12-00012.}

\dataavailability{The code used for modeling is accessible on request. Please, contact the first author.}


\conflictsofinterest{The authors declare no conflicts of interest.} 


\abbreviations{Abbreviations}{
The following abbreviations are used in this manuscript:\\

\noindent 
\begin{tabular}{@{}ll}
AXP & Anomalous X-ray pulsar \\
NS & Neutron star\\
SGR & Soft gamma-ray repeater \\
\end{tabular}
}

\appendixtitles{no} 

\begin{adjustwidth}{-\extralength}{0cm}
\reftitle{References}

\externalbibliography{yes}
\bibliography{references}

\PublishersNote{}
\end{adjustwidth}
\end{document}